\documentclass[journal]{IAENGtran}
\usepackage{bm}
\usepackage{amsfonts, amssymb}

\ifCLASSINFOpdf
   \usepackage[pdftex]{graphicx}
   \DeclareGraphicsExtensions{.pdf,.jpeg,.png}
\else
   \usepackage[dvips]{graphicx}
   \DeclareGraphicsExtensions{.eps}
\fi
\usepackage[cmex10]{amsmath}

\begin{document}
%
\title{Generating Very-High-Precision Frobenius Series\\
with Apriori Estimates of Coefficients
}
%
%
%

\author{Amna Noreen and K{\aa}re~Olaussen
 \thanks{Manuscript received August 22, 2012.}
 \thanks{A.~Noreen is with the Department
 of Physics, NTNU, N-7048 Trondheim, Norway.
 e-mail: Amna.Noreen@ntnu.no.}
 \thanks{K.~Olaussen is with the Department
 of Physics, NTNU, N-7048 Trondheim, Norway.
 e-mail: Kare.Olaussen@ntnu.no.}}

\maketitle

\pagestyle{empty}
\thispagestyle{empty}

\begin{abstract}
The Frobenius method can be used to compute solutions
of ordinary linear differential equations by generalized power
series. Each series converges in a circle which at least extends
to the nearest  singular point; hence exponentially fast inside
the circle. This makes this method well suited for very-high-precision
solutions of such equations. It is useful for this purpose to have
prior knowledge of the behaviour of the series.  We show that 
the magnitude of its coefficients can be apriori predicted to surprisingly
high accuracy, employing a Legendre transformation of the
WKB approximated solutions of the equation.
\end{abstract}

\begin{IAENGkeywords}
Second order ODEs, Regular singular points, Frobenius method,
Legendre transformation, WKB approximation.
\end{IAENGkeywords}

%
\IAENGpeerreviewmaketitle

\section{Introduction}

\IAENGPARstart{A}{ large} set of problems from technology
and science involves the study of linear second order ordinary
differential equations. Although the original problem
is more likely to be a partial differential equation
involving the Laplace operator in two or higher dimensions,
it is often possible to reduce it to a set of one-dimensional
problems through separation of variables.

For example, in three dimensions the wave, heat, or
Schr{\"o}dinger equation in zero or constant potential
can be separated in ellipsoidal coordinates $\left(\xi_1, \xi_2, \xi_3 \right)$,
related to cartesian coordinates by
\begin{align}
    x &= \sqrt\frac{(\xi_1^2-a^2)(\xi_2^2-a^2)(\xi_3^2-a^2)}{a^2(a^2-b^2)},\nonumber\\
    y &= \sqrt\frac{(\xi_1^2-b^2)(\xi_2^2-b^2)(\xi_3^2-b^2)}{b^2(b^2-a^2)},\\
    z &= \frac{\xi_1 \xi_2 \xi_3}{ab},\nonumber
\end{align} 
where $\xi_1 > a > \xi_2 > b > \xi_3 > 0$,
plus 10 degenerate forms of these coordinates~\cite{MorseFeshbach}.
The separated equations have five regular singular points,
at $\pm a$, $\pm b$, and $\infty$~\cite{MorseFeshbach2}. This
means that the equations can be formulated as second order
equations with polynomial coefficients. 
The majority of the special functions of mathematical physics,
as f.i. discussed by Whittaker and Watson \cite{WhittakerWatson}, can be
described as solutions to such equations.

The Schr{\"o}dinger equation remain separable if we add a potential of the form
\begin{equation}
    V = \frac{(\xi_2^2-\xi_3^2)u(\xi_1)+(\xi_1^2-\xi_3^2)v(\xi_2)+(\xi_1^2-\xi_2^2)w(\xi_3)}{
      (\xi_1^2-\xi_2^2)(\xi_1^2-\xi_3^2)(\xi_2^2-\xi_3^2)}.
\end{equation}
The degenerate forms often lead to situations where two or more regular singular points
merge to irregular singular points (confluent singularities).

It is sometimes useful to evaluate solutions to much higher accuracy
than possible with standard double-precision methods.
Recently we have developed and used code for solving
a large class of ordinary Frobenius type equations
to almost arbitrary high precision, in a number of algebraic
operations $N$ which grows asymptotically linearly with the
desired precision $P$, i.e.
\begin{equation}
   N \approx N_0 + \text{const}\, P.
\end{equation}
It was f.i.~used in \cite{CPC2011:AsifAmnaKareIngjald}
to find the lowest eigenvalue of
\begin{equation}
   -\psi''(x) + x^4 \psi(x) = \varepsilon \psi(x) 
\end{equation}
to an accuracy of one million decimals digits,
and its eigenvalue number $50\,000$ to 
$50\,000$ decimal digits. It was further demonstrated
in \cite{ICCP2011:AmnaKare} that it is possible to
compute normalization integrals of the resulting
wavefunctions to comparable precision. The method
also extends to computation of many other types
of amplitudes. 

In reference~\cite{CPC2012:AmnaKare}  we have published and
tested code for solving equations of the class

{\footnotesize
\begin{equation}
     -\!\left(\frac{d^2}{dz^2}  + 
     \frac{1\!-\!\nu_+ \!-\! \nu_-}{z}\frac{d}{dz} + 
     \frac{\nu_+ \nu_-}{z^2} \right)\psi(z) + 
   \frac{1}{z}\sum_{n=0}^{N} \text{v}_n\, z^n\,\psi(z) = 0.
   \label{ODE}
\end{equation}
}

\noindent
I.e., equations with polynomial coefficients (after multiplication by
$z^ 2$), and at most one regular singular point in the finite plane.

When implementing the Frobenius method \cite{FrobeniusMethod}
numerically the solution is represented by a convergent series
\begin{equation}
  \psi(z) = \sum_{m=0}^\infty a_m\,z^{m+\nu},
  \label{PowerSeries}
\end{equation}
where the coefficients $a_m$ is generated recursively in parallel
with the accumulation of the sum of the series. The method is straightforward
to extend to the  wider class of equations~(\ref{SecondOrderOde}),
but this requires consideration of several special cases.

The individual terms 
in (\ref{PowerSeries}) may grow very big, leading to huge cancellations
and large roundoff errors. It is therefore useful to have some prior
knowledge of the magnitude of the $a_m$'s before a 
high-precision evaluation --- to set the computational precision
required for a desired accuracy of the final result, and to
estimate the time required to complete the computation.
There is also the question whether one should evaluate $\psi(z)$
directly at some far away point $z$, or if it is better to make
one or more steps of analytic continuation. I.e., evaluate
$\psi(z_i)$ and $\psi'(z_i)$ at one or more intermediate
points $z_i$. Analytic continuation of functions which
satisfy a second order differential equation is rather
simple to implement, since the function is fully specified
by just two complex numbers $\psi(z_i)$ and $\psi'(z_i)$,
plus the differential equation.

We have found that $\vert a_m\vert$ can be
estimated surprisingly accurate from a WKB approximation
of the solution, followed by a Legendre transform with
additional corrections. For the general class of equations~(\ref{ODE}),
or its extension to (\ref{SecondOrderOde}), the WKB integrals
and the Legendre transform must be done by numerically, but for
this one can employ standard double-precision methods. 
The rest of this paper is organized as follows:

In section~\ref{Sec:FrobeniusMethod} we give a systematic
presentation of all explicit formulas for a Frobenius series solutions
of equation~(\ref{SecondOrderOde}) around ordinary
and regular singular points, also considering the
special cases where logarithmic terms will or may occur.

In section~\ref{Sec:LegendreTransformation} we first give a
brief motivation of the method of Legendre transformations,
from the point of view of evaluating partition function integrals
of statistical physics. We next show how the method may 
be improved by a ``finite size correction'', and demonstrate
how it works on an example with known result. Although the
correction method was motivated by our desire to estimate
the coefficients of~(\ref{PowerSeries}) more accurately, it should
also be applicable to the statistical mechanics of small systems
(i.e., systems which must be considered before the
thermodynamic limit).

In section~\ref{Sec:LegendreTransformApplication} we view
the sum~(\ref{PowerSeries}) as a partition function, and
apply the method of the previous section to find a
relation between the magnitudes $\vert a_m \vert$ and
$\vert \psi(z) \vert$. This can be used to estimate 
$\vert a_m \vert$ provided we know $\vert \psi(z) \vert$.

In section~\ref{Sec:WKBApproximation} we discuss how one
may use the WKB approximation to find a sufficiently good
approximation to  $\psi(z)$, thereby
completing the set of tools required for our estimates.

In section~\ref{Sec:Examples} we demonstrate how the method works
on a set of examples where much of the calculations may be performed
analytically.

A first account of this work has been presented in \cite{London2012}.

\section{The Frobenius method for second order ODEs}\label{Sec:FrobeniusMethod}

The Frobenius method for solving homogeneous linear ordinary differential equations
is treated in all books on differential equations. However, it is difficult to find
general expressions which are sufficiently explicit for implementation as numerical
code, and which also cover all special cases. We present such expressions in this
section.

We consider the second order differential operator
\begin{equation}
    {\cal L} =  p(z)\frac{d^2}{dz^2} + q(z)\frac{d}{dz}  + r(z),
    \label{DifferentialOperator}
\end{equation}
where 
\begin{align}
  p(z) &= \sum_{k\ge0} p_k\, z^k = C\,\prod_{n\ge0} (z-z_n)\nonumber\\[-2.8ex]
  \\
  q(z) &= \sum_{k\ge0} q_k\,z^k,\quad
  r(z) = \sum_{k\ge0} r_k\,z^k,\nonumber
\end{align}
are (short) polynomials in $z$. Solutions to the equation
\begin{equation}
  {\cal L}\,\psi(z) = 0
  \label{SecondOrderOde}
\end{equation}
in the vicinity of $z=0$, depending on the behaviour of $p(z)$,
can often be found by series expansion.
For explicit implementation of solution algorithms we must
consider several cases.

\subsection{Ordinary points}

When $p_0\ne 0$ the point $z=0$ is an ordinary point for equation~(\ref{SecondOrderOde}).
The solution can be expanded in a Taylor series,
\begin{equation}
     \psi(z) = \sum_{m\ge0} a_{m}\,z^{m}.
     \label{TaylorExpansion}
\end{equation}
Insertion into equation~(\ref{SecondOrderOde}) gives
\begin{equation}
     {\cal L}\, \psi(z) = \sum_{m\ge 0} \sum_{k\ge 0} P_k(m+2)\,a_{m+2-k}\, z^{m} = 0,
     \label{PowerExpandedEquation}
\end{equation}
with
\begin{equation}
    P_k(\mu) \equiv (\mu-k)\left[(\mu-1-k) p_k + q_{k-1}\right] + r_{k-2}.
    \label{Pdefinition}
\end{equation}
Here, and in the following, all coefficients with negative indices should be interpreted as zero,
\begin{equation*}
    a_{-n} = p_{-n} = q_{-n} = r_{-n} = 0,\quad \text{for } n=1,2, \ldots.
\end{equation*}
The requirement that the coefficient
of each power $z^m$ in (\ref{PowerExpandedEquation}) must vanish
leads to the recursion formula
\begin{equation}
   a_{m+2} = -\frac{1}{(m+2)(m+1)\,p_0}\,\sum_{k\ge 1} P_k(m+2)\,a_{m+2-k}.
   \label{RecursionFormula1}
\end{equation}
The  coefficients $a_0$ and $a_1$ can be chosen freely, or according to the initial conditions.

\subsection{Regular singular points}

When (a) $p_0 = 0$ with $p_1\ne 0$ (and at least one of $ q_0$ or $r_0$ is nonzero),
or when (b) $p_0=p_1=q_0=0$ with $p_2 \ne 0$ (and $r_0$ is nonzero),
equation~(\ref{SecondOrderOde}) has a regular singular point at $z=0$.
A series solution can be found by the Frobenius method.
I.e., one writes the tentative solution as a generalized
power series (\ref{PowerSeries}).

One solvability condition is that $\nu$ must satisfy a second order
algebraic equation. Hence, counting possible degeneracies,
there will be two solutions, $\nu_{1,2}$ (choosing $\text{Re}\, \nu_1  \le \text{Re}\,\nu_2$).
We must further consider the cases (i) $\nu_1 = \nu_2$, (ii) $\nu_2 = \nu_1+\ell$ with
$\ell$ a positive integer, and (iii) {\em everything else\/}.
To avoid explicit implementation of too many special cases
we can instead write
\begin{equation}
    \psi(z) = z^{\nu_2}\,\tilde{\psi}(z),
\end{equation}
and solve the resulting equation for $\tilde{\psi}(z)$. 
The explicit transformation of the polynomial coefficients is
\begin{align}
    \tilde{p}(z) &= z^{-1}\,p(z),\nonumber\\
    \tilde{q}(z) &= 2\nu z^{-2} p(z) + z^{-1} q(z),\\
    \tilde{r}(z)  &= \nu(\nu-1) z^{-3} p(z) + \nu z^{-2} q(z) + z^{-1} r(z),\nonumber
\end{align}
where $\tilde{p}(z)$ and $\tilde{q}(z)$ remain polynomials because $p_0=p_1=q_0=0$,
and $\tilde{r}(z)$ is a polynomial when
\begin{equation}
    \nu(\nu-1) p_2 + \nu q_1 + r_0 =0.
\end{equation}
I.e., when $\nu$ satisfies the indicial equation. 
The transformed equation corresponds to the case (a) above, with index $\tilde{\nu}_2 = 0$ and
$q_0/p_1 \ge 1$. The latter implies $\tilde{\nu}_1 \le 0$.
Henceforth we will only consider this case, dropping $\tilde{\ }$ from the notation. 
For the computations below we need various cases of the formula (using $P_0(\mu)=0$ when
$p_0 =0$),
\begin{align}
    &{\cal L}\,\sum_{m\ge0} \left( a_{0,m} + a_{1, m} \log z\right) z^{m+\nu} =\nonumber\\
    &\sum_{m\ge -1} \Big\{\sum_{k\ge 0} \Big[ P_{k\!+\!1}(m\!+\!2\!+\!\nu)\left( a_{0,m\!+\!1\!-\!k} + a_{1, m\!+\!1\!-\!k} \log z\right)\nonumber\\
    &  +  Q_{k\!+\!1}(m\!+\!2\!+\!\nu)\,a_{1, m\!+\!1\!-\!k}  \Big]\Big\} z^{m+\nu},
\end{align}
with  $P_k(\mu)$ given by equation~(\ref{Pdefinition}), and
\begin{equation}
    Q_k(\mu) = (2\mu-1 -2k)p_k + q_{k-1}.
    \label{Qdefinition}
\end{equation}

\subsection{Regular singular point with non-integer index difference}

First consider
the {\em everything else\/} case. We assume a solution of the
form~(\ref{PowerSeries}). Now ${\cal L} \psi(z)=0$ implies
\begin{equation}
      \sum_{k\ge 0} P_{k+1}(m+2+\nu)\,a_{m+1-k} = 0,
     \label{SeriesExpandedEquation}
\end{equation}
for $m=-1, 0,\ldots$. For $m=-1$  this becomes
\begin{equation}
     P_{1}(1+\nu) = \nu \left[(\nu-1)p_1 + q_0 \right] = 0,
\end{equation}
with solutions and $\nu_1 = 1 - q_0/p_1$, and $\nu_2 =0$.
In this case $\nu_1$ is non-integer. For $m\ge 0$, and $\nu=\nu_1$ and $\nu=\nu_2=0$
respectively, equation~(\ref{SeriesExpandedEquation}) implies the recursion relations
\begin{align}
     a_{m\!+\!1} &= \frac{-1}{(m\!+\!1)(m \!+\! 1 \!+\! \nu_1)p_1}\sum_{k\ge 1} P_{k+1}(m\!+\!2\!+\!\nu_1)\,a_{m\!+\!1\!-\!k},
     \nonumber\\[-2.0ex]
   \label{RecursionFormula2}\\
   a_{m\!+\!1} &= \frac{-1}{(m\!+\!1)(m \!+\! 1 \!-\! \nu_1) p_1}\sum_{k\ge 1}\, P_{k+1}(m\!+\!2)\,a_{m\!+\!1\!-\!k}.
   \nonumber
\end{align}
In both cases the coefficient $a_0$ can be chosen freely,
or according to initial conditions.
The last recursion in (\ref{RecursionFormula2}) is always working,
since by arrangement $\nu_1 \le 0$. The first one
is working in this case since $\nu_1$ is non-integer by assumption. 

\subsection{Regular singular point with degenerate indices}

For the degenerate case of $\nu_1=\nu_2 = 0$ we still have a
solution corresponding to the last
recursion in equation~(\ref{RecursionFormula2}). 
For a linear independent solution we make the ansatz
\begin{equation}
    \psi(z) = \sum_{m\ge 0} \left(a_{0,m} + a_{1,m} \log z \right) z^m.
\end{equation}
Now ${\cal L}\,\psi(z)=0$ implies
\begin{align}
    &\sum_{k\ge 0} P_{k\!+\!1}(m\!+\!2)\,a_{1, m\!+\!1\!-\!k} = 0,\nonumber\\[-2.0ex]
    &\label{CoefficientConditions}\\
    &\sum_{k\ge 0} P_{k\!+\!1}(m\!+\!2)\,a_{0, m\!+\!1\!-\!k}  + Q_{k\!+\!1}(m\!+\!2) \,a_{1, m\!+\!1\!-\!k} = 0.\nonumber
\end{align}
for $m=-1, 0,\ldots$. For $m=-1$  this becomes
\begin{equation}
  P_1(1) = Q_1(1) =0, 
\end{equation}
which holds since $q_0=p_1$ when $\nu_1=0$. 
For $m\ge 0$ equation~(\ref{CoefficientConditions}) implies the recursion relations
\begin{align}
   a_{1,m\!+\!1} &= \frac{-1}{(m\!+\!1)^2 p_1}
   \sum_{k\ge 1}\, P_{k\!+\!1}(m\!+\!2)\,a_{1,m\!+\!1\!-\!k},\nonumber\\
   a_{0,m\!+\!1} &=  \frac{-1}{(m\!+\!1)^2 p_1}
   \Big[ \sum_{k\ge 1}\, P_{k\!+\!1}(m\!+\!2)\,a_{0,m\!+\!1\!-\!k} \label{RecursionFormula3}\\ 
   &\phantom{= \frac{-1}{(m\!+\!1)^2 p_1}}
   + \!\!\sum_{k\ge 0}\, Q_{k\!+\!1}(m\!+\!2)\,a_{1,m\!+\!1\!-\!k}\Big].\nonumber
\end{align}
The coefficients $a_{0,0}$ and $a_{1,0}$ can be chosen freely,
or according to initial conditions.

\subsection{Regular singular point with integer index difference}

For the case $\nu_1=-\ell$ (a negative integer) and $\nu_2 =0$ we still
have a ($\nu_2$) solution corresponding to the last recursion in
equation~(\ref{RecursionFormula2}). For a linear independent solution
we make the ansatz
\begin{equation}
    \psi(z) = \sum_{m\ge 0} \left(a_{0,m} + a_{1,m} \log z \right) z^{m-\ell}.
\end{equation}
Now ${\cal L}\,\psi(z)=0$ implies
\begin{align}
    &\sum_{k\ge 0} P_{k\!+\!1}(m\!+\!2\!-\!\ell)\,a_{1, m\!+\!1\!-\!k} = 0,\nonumber\\[-2.0ex]
    &\label{CoefficientConditions2}\\
    &\sum_{k\ge 0} P_{k\!+\!1}(m\!+\!2\!-\!\ell)\,a_{0, m\!+\!1\!-\!k}  \!+\! Q_{k\!+\!1}(m\!+\!2\!-\!\ell) \,a_{1, m\!+\!1\!-\!k} = 0.\nonumber
\end{align}
for $m=-1, 0,\ldots$. For $m=-1$  this can be solved by chosing $a_{0,0}$ arbitrary, 
since $P_1(1-\ell) = 0$ when $\nu_1 = -\ell$ and $q_0 =(1+\ell) p_1$,
and $a_{1,0} = 0$ since $Q_1(1-\ell) =-\ell p_1 \ne 0$. We may next use the first recursion in
equation~(\ref{RecursionFormula2}) to compute $a_{0,1+m}$ until $m+1 = \ell$, 
where it breaks down. For $m+1 = \ell$ equation~(\ref{CoefficientConditions2}) becomes
\begin{equation}
     \sum_{k\ge 1} P_{k+1}(1) a_{0,\ell-k} + Q_{1}(1) a_{1,\ell} = 0,
\end{equation}
with solution (note that $Q_1(1) = \ell p_1$)
\begin{equation}
       a_{1,\ell} = -\frac{1}{\ell p_1} \sum_{k\ge 1} P_{k+1}(1) a_{0,\ell-k}. 
\end{equation}
We may choose $a_{0,\ell}$ freely; 
this correponds to an addition of the $\nu_2$ solution. 
For $m+1 \ge \ell$ equation~(\ref{CoefficientConditions2}) implies the recursion relations
\begin{align}
   a_{1,m\!+\!1} &= \frac{-1}{(m\!+\!1)(m\!+\!1\!-\!\ell) p_1}
   \sum_{k\ge 1} P_{k\!+\!1}(m\!+\!2\!-\!\ell)\,a_{1,m\!+\!1\!-\!k},\nonumber\\[-2ex]
   \label{RecursionFormula4}\\
   a_{0,m\!+\!1} &=  \frac{-1}{(m\!+\!1)(m\!+\!1\!+\!\ell) p_1}
   \Big[\!\sum_{k\ge 1} P_{k\!+\!1}(m\!+\!2\!-\!\ell)\,a_{0,m\!+\!1\!-\!k}\nonumber \\ 
   &\phantom{= \frac{-1}{(m\!+\!1)(m\!+\!1\!+\!\ell) p_1}}
   \!\!\!\!\!\!\!\! + \sum_{k\ge 0} Q_{k\!+\!1}(m\!+\!2\!-\!\ell)\,a_{1,m\!+\!1\!-\!k}\Big].\nonumber
\end{align}
This completes the description of the Frobenius series solution of
${\cal L}\, \psi(z) = 0$ around ordinary and regular singular points.

\section{Legendre transform with finite size corrections}\label{Sec:LegendreTransformation}

Although the Legendre transform is a well-defined mathematical
procedure by itself \cite{LegendreTransform1, LegendreTransform2, LegendreTransform3}, 
it can f.i.~be motivated as a maximum integrand
approximation to integral relations between partition functions
in equilibrium statistical physics. As such it can be modified
to take into account corrections due to (gaussian) integral
contributions around the maximum.

\subsection{Maximum integrand approximation}

Consider the integral relation
\begin{equation}
    \text{e}^{F(\bm{p})} =  
    \int \text{d}\bm{y}\,\text{e}^{-U(\bm{y}) + \bm{p}\cdot\bm{y}},
    \label{IntegralTransform}
\end{equation}
where $U(\bm{y})$ is assumed to be a smooth function of
its argument. We first approximate the integral by the maximum value
of the integrand. This occur at some point $\bm{x}$ where
\begin{equation*}
    -\bm{\nabla}_{\bm{x}}U(\bm{x}) + \bm{p} = 0,
\end{equation*}
and the approximate value of $F(\bm{p})$, which we shall denote $F_0(\bm{p})$,
becomes $\bm{p}\cdot\bm{x} - U(\bm{x})$. These relations
correspond to the standard Legendre transform from the
pair $\bm{x}$, $U(\bm{x})$ to
\begin{align}
     \bm{p} &= \bm{\nabla}_{\bm{x}}{U}(\bm{x}),\label{IndependentVariables}\\
     F_0(\bm{p}) &= \bm{p}\cdot\bm{x}-U(\bm{x}),\label{DependentQuantity}
\end{align}
where $\bm{x}$ is assumed to be eliminated by
use of equation~(\ref{IndependentVariables}).

Consider next the effect of changing $\bm{p}$ by a
small amount, $\bm{p} \to \bm{p} + \bm{\delta p}$.
This will change the maximum $\bm{x}$ by a
small amount, $\bm{x} \to \bm{x} + \bm{\delta x}$.
This leads to a small change in $F_0$,
\begin{align*}
    F_0(\bm{p}) &\to  F_0(\bm{p} + \bm{\delta p}) = F_0(\bm{p}) + \bm{\delta p}\cdot \bm{\nabla}_{\bm{p}}F_0(\bm{p})\\
    &= \left(\bm{p} + \bm{\delta p} \right)\cdot\left(\bm{x} + \bm{\delta x} \right) - U(\bm{x} + \bm{\delta x})\\
    &= \bm{p}\cdot\bm{x} - U(\bm{x}) + \bm{\delta x}\cdot\left[\bm{p} - \bm{\nabla}_{\bm{x}} U(\bm{x})\right] + \bm{\delta p}\cdot \bm{x},
\end{align*}
when neglecting second order corrections. Eliminating terms by use of (\ref{IndependentVariables}) and (\ref{DependentQuantity})
gives $\bm{\delta p}\cdot \bm{\nabla}_{\bm{p}}F_0(\bm{p}) = \bm{\delta p}\cdot \bm{x}$, or since the direction of $\bm{\delta p}$ is arbitrary,
\begin{align}
      \bm{x} &= \bm{\nabla}_{\bm{p}} F_0(\bm{p}),\label{IndependentVariables2}\\
     U(\bm{x}) &= \bm{p}\cdot\bm{x}-F_0(\bm{p}).\label{DependentQuantity2}   
\end{align}
This is the inverse Legendre transform. Equation~(\ref{DependentQuantity2}) is
just a trivial rewriting of (\ref{DependentQuantity}).

It follows from equation (\ref{IndependentVariables}) that the Jacobi matrix
\begin{equation*}
     \left(\frac{\partial p^i}{\partial x^j} \right) = \frac{\partial^2}{\partial x^j \partial x^i} U(\bm{x}),
\end{equation*}
and from equation (\ref{IndependentVariables2}) that the Jacobi matrix
\begin{equation*}
   \left(\frac{\partial x^i}{\partial p^j} \right) = \frac{\partial^2}{\partial p^j \partial p^i} F_0(\bm{p}).
\end{equation*}
Since Jacobi matrices of inverse transformations are matrix inverses, we have the relation
\begin{equation}
    U_{ij}(\bm{x}) \equiv \left(\frac{\partial^2  U(\bm{x})}{\partial x^i \partial x^j}\right) = 
    \left(\frac{\partial^2  F_0(\bm{p})}{\partial p^i \partial p^j} \right)^{-1} \equiv F^{-1}_{0\,ij}(\bm{p}).
    \label{InverseMatrixRelation}
\end{equation}

\subsection{One loop fluctuation correction}

We next improve the evaluation of (\ref{IntegralTransform}) by expanding $U(\bm{y})$
around its minimum. We write $\bm{y} = \bm{x} + \bm{z}$ and assume that the main
contribution to the integral comes from a small range of $\bm{z}$-values.  Hence
\begin{equation*}
    U(\bm{x}+\bm{z}) \approx U(\bm{x}) + \bm{z}\cdot\bm{\nabla}_{\bm{x}} U(\bm{x}) + 
    \frac{1}{2}z^i z^j U_{ij}(\bm{x}),
\end{equation*}
which gives
\begin{align}
    \text{e}^{F(\bm{p})}  &= \text{e}^{\bm{p}\cdot\bm{x} - U(\bm{x})}
    \int \text{d}\bm{z}\,\text{e}^{-\frac{1}{2} z^i z^j U_{ij}(\bm{x})}\nonumber\\
    &= \text{e}^{\bm{p}\cdot\bm{x} - U(\bm{x})}\,\det\left(U_{ij}(\bm{x})/2\pi\right)^{-1/2}.
\end{align}
I.e., denoting the corrected expression $F_1(\bm{p})$,
\begin{align}
     F_1(\bm{p}) &= \bm{p}\cdot \bm{x} - U(\bm{x}) - 
     \frac{1}{2} \log \det\left( U_{ij}(\bm{x})/2\pi\right)\nonumber\\
     &\equiv F_0(\bm{p}) - \frac{1}{2}\log \det\left( U_{ij}(\bm{x})/2\pi\right).
     \label{OneLoopCorrection}
\end{align}
This corresponds to a standard one-loop correction to the partition function
$\text{e}^{F(\bm{p})}$ of statistical physics, where one may proceed with an
ordinary inverse Legendre transform to define a (fluctuation corrected)
{\em effective potential\/}
\begin{align}
      \bm{x} &= \bm{\nabla}_{\bm{p}} F_1(\bm{p}),\label{IndependentVariables3}\\
      \Gamma(\bm{x}) &= \bm{p}\cdot\bm{x} - F_1(\bm{p}).\label{DependentQuantity3}
\end{align} 
However, here we want to compute the original
potential $U(\bm{x})$ from the information provided
by $F_1(\bm{p})$. The relation~(\ref{InverseMatrixRelation}) now reads
\begin{equation}
     U_{ij}(\bm{x}) = \left(\frac{\partial^2  F_0(\bm{p})}{\partial p^i \partial p^j} \right)^{-1},
\end{equation}
but we have no direct access to the quantity $F_0(\bm{p})$. Equation~(\ref{OneLoopCorrection})
instead becomes a second order differential equation relating $F_0(\bm{p})$; hence an exact
inversion seems difficult in general. However, if the one-loop correction is assumed to be small
compared to $F_0(\bm{p})$, we may approximate
\begin{equation*}
  \left(\frac{\partial^2  F_0(\bm{p})}{\partial p^i \partial p^j} \right) 
  \approx \left(\frac{\partial^2  F_1(\bm{p})}{\partial p^i \partial p^j} \right) \equiv F_{1\,ij}(\bm{p}),
\end{equation*}
and
\begin{equation}
     F_0(\bm{p}) \approx F_1(\bm{p}) - \frac{1}{2}\log\det \left( 2\pi F_{1\,ij}(\bm{p})\right).
\end{equation}
We finally apply the inverse Legendre transform 
(\ref{IndependentVariables2}), (\ref{DependentQuantity2}) of the
pair $\bm{p}, F_0(\bm{p})$ to find the pair $\bm{x}, U(\bm{x})$.

\subsection{Simple application}

\begin{figure}[!t]
\begin{center}
\includegraphics[clip, trim = 8ex 6ex 9ex 5ex, width=0.483\textwidth]{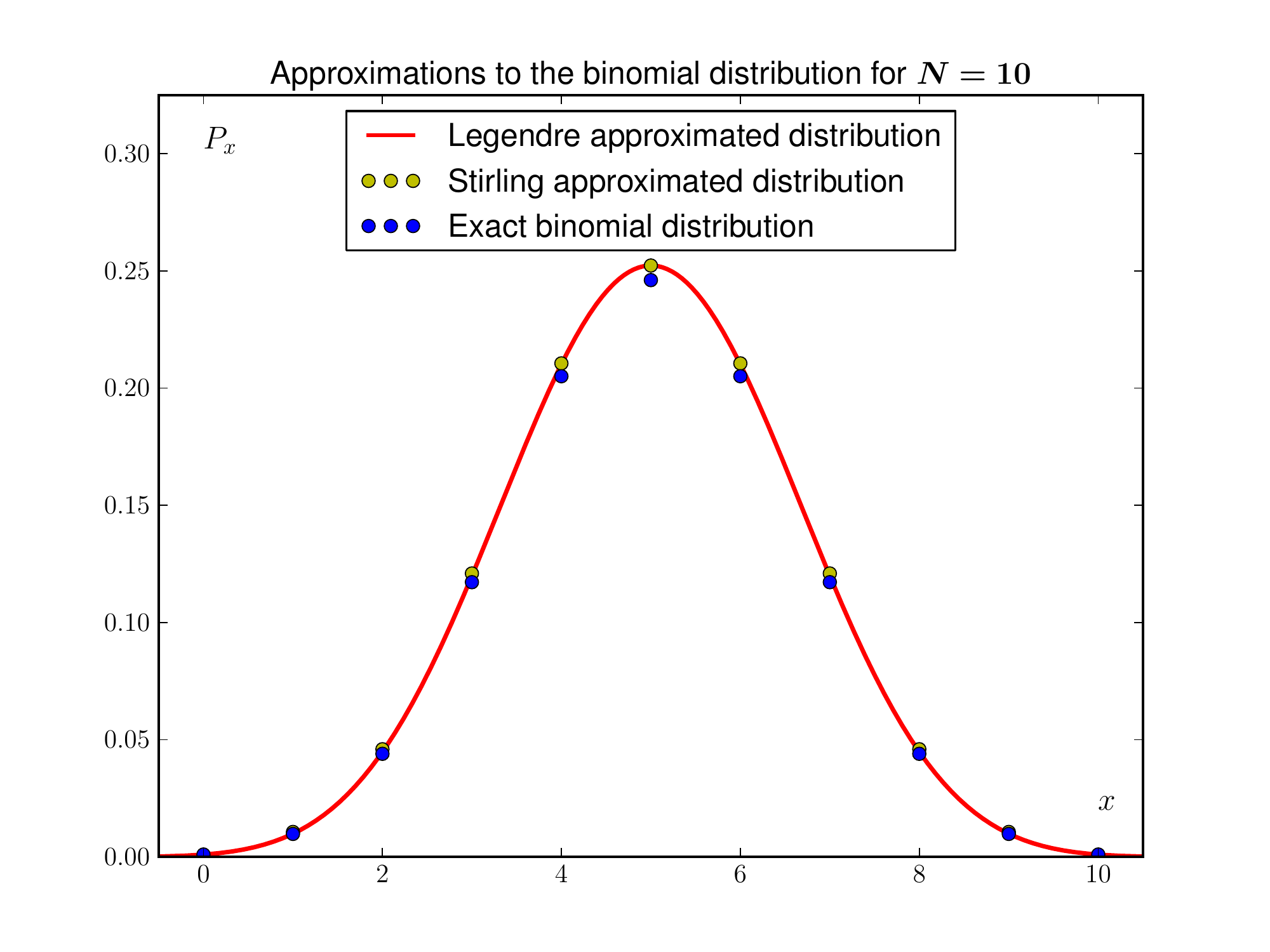}
\end{center}
\caption{Comparison of the continuous distribution obtained by a
Legendre transform (with finite size corrections)
of the generating function, cf.~equation~(\ref{LegendreApproximatedPx}),
with the exact binomial distribution (\ref{ExactProbability}) for $N=10$.
For a fair comparison one should perhaps compare the exact probability $P_x$
given by (\ref{ExactProbability}) with the average value of (\ref{LegendreApproximatedPx})
over the interval $\left(x-\frac{1}{2}, x+\frac{1}{2}\right)$.
The result of making a Stirling approximation of the binomial coefficient
is also shown. This approximation gives $P_x = 0$ at the endpoints
$x=0$ and $x=N$.
}
\label{Comparison10}
\end{figure}

\begin{figure}[!t]
\begin{center}
\includegraphics[clip, trim = 8ex 6ex 9ex 5ex, width=0.483\textwidth]{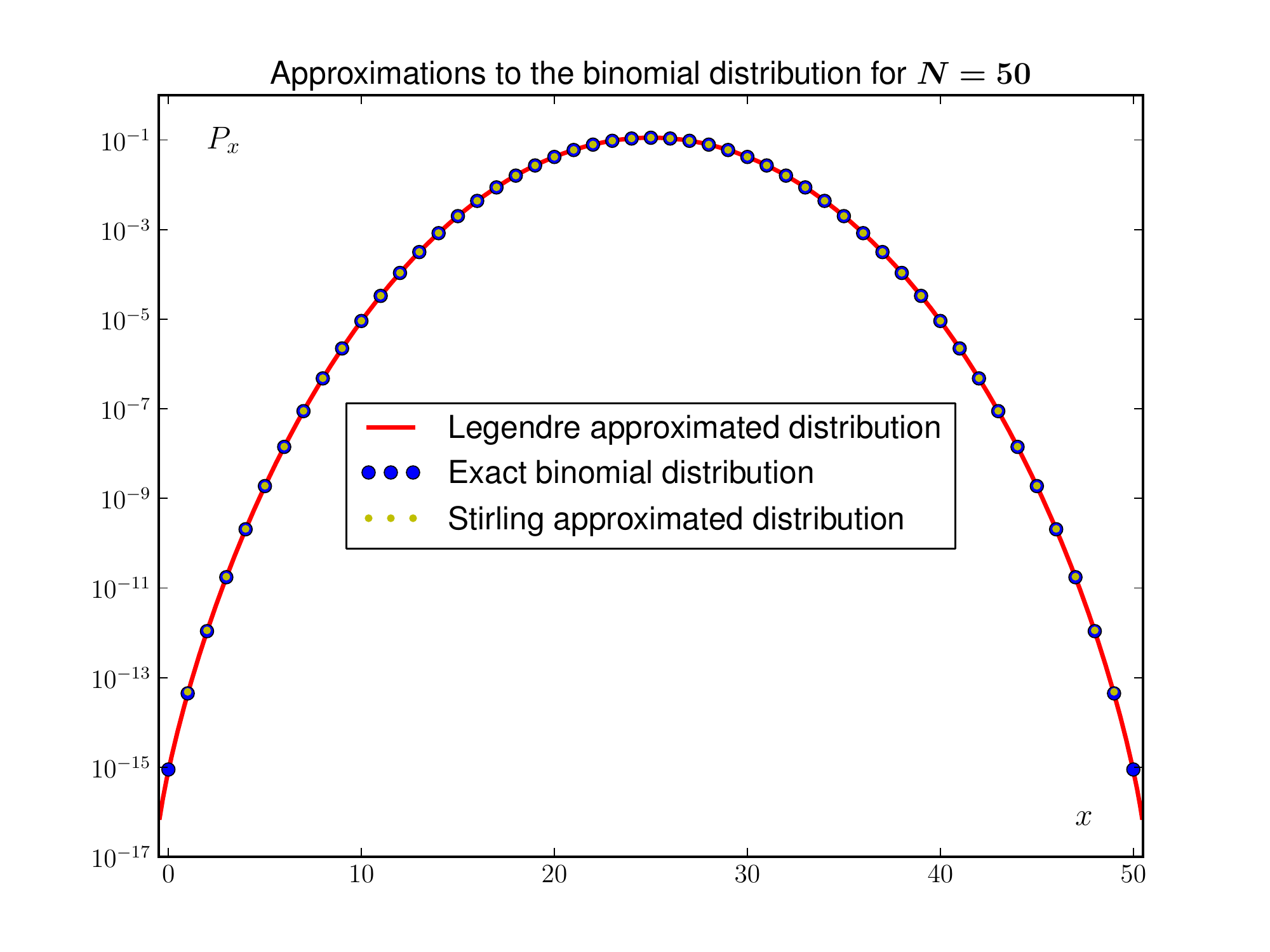}
\end{center}
\caption{Comparison of the continuous distribution obtained by a
Legendre transform (with finite size corrections)
of the generating function, cf.~equation~(\ref{LegendreApproximatedPx}),
with the exact binomial distribution (\ref{ExactProbability}) for $N=50$.
The result of making a Stirling approximation of the binomial coefficient
is also shown. This approximation gives $P_x = 0$ at the endpoints
$x=0$ and $x=N$.
}
\label{Comparison50}
\end{figure}

Consider the  ``partition function''
\begin{equation}
      \text{e}^{f(p)} = 2^{-N}\,\left(1+\text{e}^p\right)^N.
\end{equation}
This actually is the generating function for the probability of
$x$ heads in a sequence of $N$ independently flipped coins,
corresponding to the probability
\begin{equation}
    P_x \equiv \text{e}^{-u(x)} = 2^{-N}\,\binom{N}{x},
    \label{ExactProbability}
\end{equation}
but assume we don't know that. We instead use the method above,
with
\begin{align}
  f(p)  &=  N \log \left(1 + \text{e}^p \right) - N \log 2,\\
  f''(p) &= N\,\text{e}^p(1 + \text{e}^p)^{-2}.
\end{align}
From
\begin{align*}
    x &= \frac{d}{dp} \left[ f(p) - \frac{1}{2}\log 2\pi f''(p)\right],\\
    u(x) &= p x -  \left[ f(p) - \frac{1}{2}\log 2\pi f''(p)\right],
\end{align*}
we find
\begin{equation*}
    p = \log \xi - \log (1-\xi),
\end{equation*}
where we have written $x= (N+1)\xi - \frac{1}{2}$
to simplify expressions, and
\begin{align}
  u(x) &= (N+1) \left[\xi\log\xi + (1-\xi)\log(1-\xi)\right]\nonumber\\ 
       &+ {\textstyle\frac{1}{2}}\log 2\pi N + N \log 2.
\end{align}
This means that the discrete probability $P_x$ of equation~(\ref{ExactProbability}) 
is approximated by a continuous distribution
\begin{equation}
    \text{e}^{-u(x)} = 2^{-N}\frac{1}{\sqrt{2\pi N}}\,
    \text{e}^{-(N+1)\left[\xi\log\xi + (1-\xi)\log(1-\xi)\right]},
    \label{LegendreApproximatedPx}
\end{equation}
which should be compared with the Stirling approximation to the binomial coefficient
\begin{align}
  2^{-N}\,&\binom{N}{x} \approx 2^{-N}\frac{1}{\sqrt{2\pi N}}\nonumber\\
  &\times \frac{1}{\sqrt{\chi(1-\chi)}}
  \text{e}^{-N\left[\chi\log\chi + (1-\chi)\log(1-\chi)\right]},
    \label{StirlingApproximatedPx}
\end{align}
where $x=N\chi$. Comparisons between the exact distribution~(\ref{ExactProbability}),
its Stirling approximation~(\ref{StirlingApproximatedPx}), and
the Legendre transformed approximation~(\ref{LegendreApproximatedPx}) are shown in
Fig.~\ref{Comparison10} and \ref{Comparison50}.

\section{Estimating Frobenius coefficients by Legendre transform}\label{Sec:LegendreTransformApplication}

We now use the results of the previous section
for our main objective, to estimate the magnitude of the coefficients $a_m$.
Here we will not consider the presence of logarithmic terms. 
Our basic hypothesis is that the sum~(\ref{PowerSeries}) for large $\vert z \vert$ 
receives its main contribution from a
relatively small range of $m$-values, at least for some phase values of $z$.
Introduce quantities $u$ and $s(m)$ so that
\begin{equation*}
  x = \text{e}^u,\quad \vert a_m \vert = \text{e}^{s(m)}.
\end{equation*}
Hence our hypothesis is that
\begin{equation}
   \text{e}^{S(u)} \equiv \max_{\varphi} \psi(\text{e}^{u+i\varphi}) \approx 
   \sum_{m} \text{e}^{s(m)+(\nu+m)u},
   \label{MaximumAbsoluteValue}
\end{equation}
with the main contribution to the sum coming from a relatively small range
of $m$-values around a maximum value $\bar{m}$. The latter is
defined so that $s'(\bar{m})+u=0$, $s''(\bar{m})<0$.
Now write $m = \bar{m} + \Delta m$, and approximate the
sum (\ref{MaximumAbsoluteValue}) over $\Delta m$ by a gaussian integral.
This gives
\begin{equation*}
   \text{e}^{S(u)} \approx   \sqrt{{-2\pi}/{s''(\bar{m})}}\,\text{e}^{s(\bar{m})+(\nu+\bar{m})u}.
\end{equation*}
In summary, we have found the relations
\begin{align}
    u    &= -s'(m)\label{VariableChange},\\
    S(u) &= s(m) - (\nu + m) s'(m) +\frac{1}{2}\log\left(\frac{2\pi}{-s''(m)} \right)\nonumber\\
         &\equiv S_0(u) + \frac{1}{2}\log\left(\frac{2\pi}{-s''(m)} \right).\label{FunctionChange}
\end{align}
Compared with the discussion in the previous section we see that
this is essentially a Legendre transformation between $s(m)$ and $S(u)$,
but since there are some minor differences we repeat the derivation of
the inverse transformation.

Consider a small change $u\to u+\delta u $. To maintain the maximum condition
we must also make a small change $m \to m+\delta m$, with
\(
    \delta m = -{\delta u}/{s''(m)}
\). I.e.~$s''(m)=-u'(m)$.
This is consistent with the result of taking the $m$-derivative of
equation~(\ref{VariableChange}).
One further finds that $S_0(u)$ becomes
\begin{align*}
    &S_0(u+\delta u) = S_0(u) + S'_0(u)\,\delta u + \frac{1}{2}S''_0(u)\,\delta u^2 +\cdots\\
    &=s(m) + (\nu + m) u + (m+\nu)\,\delta u - \frac{1}{2s''(m)}\, \delta u^2 + \cdots,
\end{align*}
giving the relations
\begin{align}
    (m+\nu) &= S'_0(u), \label{mExpression}\\
    s(m)    &=  S_0(u) - u S'_0(u),\label{sExpression}\\
    s''(m)  & = -S''_0(u)^{-1}. \label{SecondDerivative}
\end{align}
Equation (\ref{SecondDerivative}) just says that $\left(dm/du\right) = \left(du/dm\right)^{-1}$.
We are only able to compute $S(u)$ directly, not $S_0(u)$. However, they only differ by
a logarithmic term, hence we will approximate $\log(-s''(m)) = -\log S''_0(u) \approx -\log S''(u)$.
This gives
\begin{equation}
    S_0(u) \approx S(u) - \frac{1}{2}\log\left(2\pi\, S''(u)\right),\label{logCorrectedS0}
\end{equation}
which can be used in equations (\ref{mExpression}--\ref{SecondDerivative})
when we have computed $S(u)$.

\section{WKB approximation}\label{Sec:WKBApproximation}

It remains to find $S(u)$. Here we will use the leading order WKB 
approximation to find a sufficiently accurate estimate.
When $z=0$ is an ordinary point, i.e. when $\nu_-=0$, $\nu_+=1$, 
the leading order WKB solution to (\ref{ODE}) is \cite{Schiff, Kroemer, BenderOrszag}
 \begin{equation}
   \psi(z) \approx \sqrt{Q_0/Q(z)}
   \exp\left({\frac{1}{s}}\int_0^{z} Q(t) \text{d}t \right),
   \label{WKBApproximation}
\end{equation}
where $Q^2(z) = \sum_{n=1}^N \text{v}_n z^{n-1}$, and $Q_0=Q(0)$.
This represents a superposition of the solutions $\psi_\pm(z)$. 
The difference between the $\nu_+$ and $\nu_-$ solutions is at worst
comparable to accuracy of our approximation; hence we will not
distinguish between them.

When $z=0$ is a regular singular point we use the Langer corrected
WKB approximation to obtain leading order solutions in the form
\begin{align}
  \psi_{\pm}(z) &\approx
  z^{\nu_{\pm}} \sqrt{Q_0/Q(z)}\; \times\nonumber\\
  &\exp\left(\pm\frac{1}{s}\int_0^z
    \frac{\text{d}t}{t} 
    \left[\sqrt{ Q^2(t)} - Q_0\right] \right). \label{LangerCorrectedWKBApproximation}
\end{align}
Here $Q^2(z) = \frac{1}{4}s^2 (\nu_+-\nu_-)^2 + \sum_{n=0}^N \text{v}_n z^{n+1}$,
and $Q_0 = Q(0)$. 
In equation (\ref{LangerCorrectedWKBApproximation}) we distinguish
between the $\nu_+$- and $\nu_-$-solutions, because the difference $\nu_+ - \nu_-$ may
in principle be large. 

The WKB integrals must in general be done numerically, sometimes along curves
in the complex plane. This requires careful attention to branch cuts. One must
also take into account that the behaviour of the WKB solution may change from one exponential behaviour
to another, due to the existence of Stokes lines \cite{FurryNotesOnPhaseIntegralMethod}.
We have observed that this sometimes can be explained as contributions from
topologically different integrations paths.
In this paper we will only give some examples where most of the calculations can be done
analytically.

\section{Examples}\label{Sec:Examples}

\subsection{Anharmonic oscillators}

Consider the equation
\begin{equation}
   -\frac{\partial^2}{\partial y^2}\Psi(y) + \left(y^2+ c^2\right)^2\Psi(y) = 0,
   \label{Anharmonic_oscillator}
\end{equation}
for  real $c$ so that $c^2 \ge 0$. For large $y$ the typical solution behaves like
\begin{equation}
     \Psi(y) \sim \text{e}^{\frac{1}{3}y^3 + c^2 y},
     \label{CrudeWKBApproximation}
\end{equation}
neglecting the slowly varying prefactor. For a given value of $\vert y \vert$
this is maximum along the positive real axis.
Hence, with $x=y^2=\text{e}^{u}$, we find as a leading approximation
\begin{equation*}
     S(u) = {\textstyle \frac{1}{3}}\left(\text{e}^{\frac{3}{2}u} + 3 c^2 \text{e}^{\frac{1}{2}u}\right).
\end{equation*}
In this case the Frobenius series can be written
\begin{equation}
  \Psi(y) = \sum_{m=0}^{\infty} a_m\, y^{2m + \nu} \equiv \sum_{m=0}^{\infty} A_m(y),
\end{equation}
with $\nu=0,\;1$.
Ignoring the $\log(S''(u))$-term in~(\ref{mExpression}, \ref{sExpression}, \ref{logCorrectedS0})
we find
\begin{align}
  {m} &= {\textstyle \frac{1}{2}}\left(\text{e}^{\frac{3}{2}u} + c^2\,\text{e}^{\frac{1}{2}u} \right),
  \label{m_Anharmonic}
  \\
  \log\left(\left| a_{{m}}\right|\right) &= 
  \left({\textstyle \frac{1}{3}} - {\textstyle \frac{1}{2}} u\right) \text{e}^{\frac{3}{2}u} + 
  c^2 \left(1 -{\textstyle \frac{1}{2}} u \right) \text{e}^{\frac{1}{2}u}.
  \label{a_m_Anharmonic}
\end{align}

\begin{figure}[!t]
\begin{center}
\includegraphics[clip, trim = 8ex 6ex 9ex 5ex, width=0.483\textwidth]{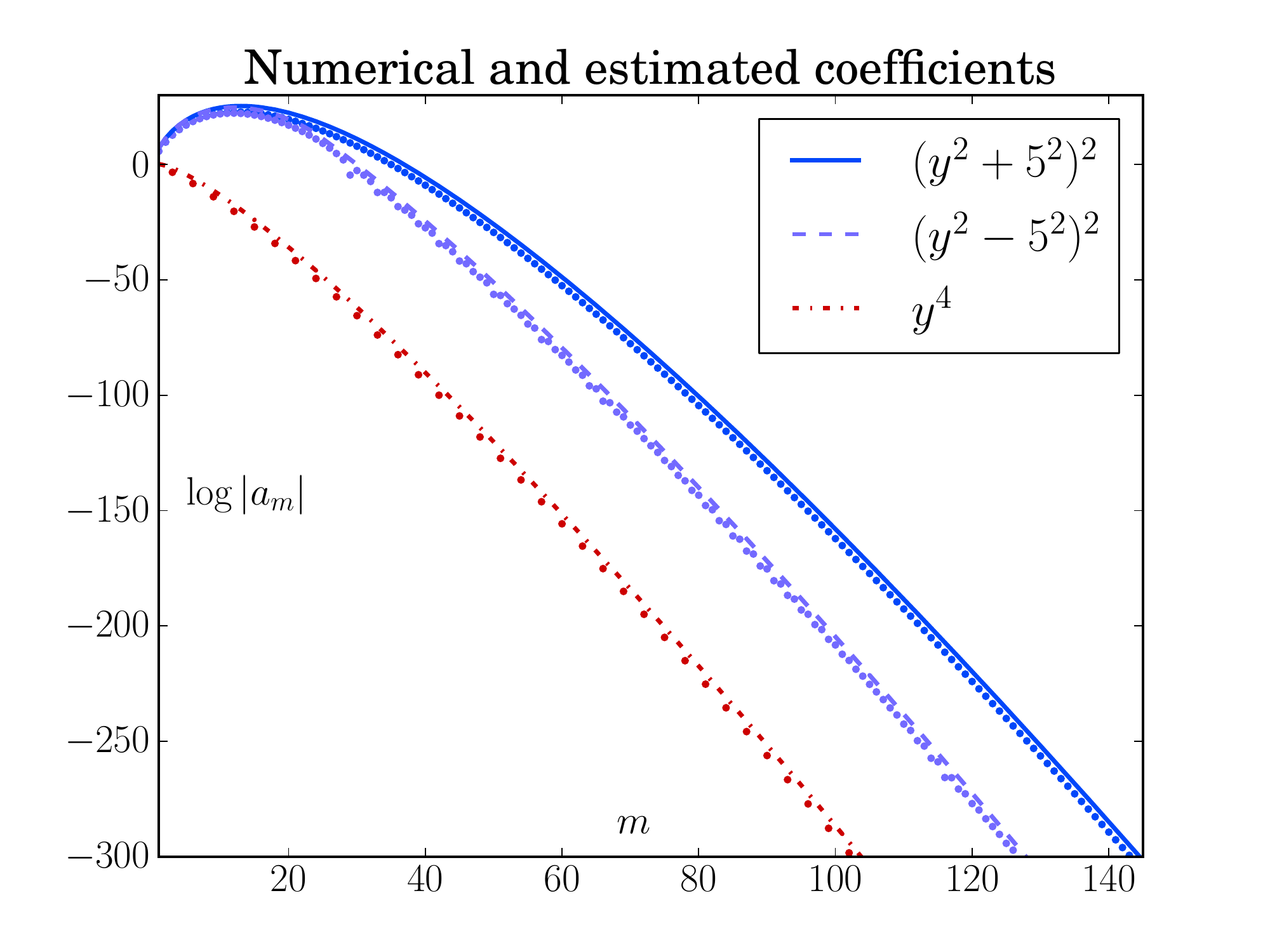}
\end{center}
\caption{Comparison of numerical coefficients $a_m$ (points)
with estimates (full-drawn lines) based on
(\ref{m_Anharmonic}, \ref{a_m_Anharmonic}) and
(\ref{m_DoubleWell}). The estimates
of $\log\vert a_m \vert$ are accurate up to corrections which
depend logarithmically on $m$.
}
\label{Coefficients_a_m}
\end{figure}

\noindent
For $c=0$ an explicit representation is
\begin{equation}
   \log \vert a_m \vert = \frac{2}{3}m\left(1 -\log 2m \right).
   \label{prediction0}
\end{equation}
This is plotted as the lower curve in figure~\ref{Coefficients_a_m}. It fits satisfactory
with the high-precision coefficients generated numerically, but there remains a
correction which depends logarithmically on $m$. For nonzero $c$ the parametric
representation provides equally good results, as shown by the upper curve in 
figure~\ref{Coefficients_a_m}.

\begin{figure}[!t]
\begin{center}
\includegraphics[clip, trim = 10.5ex 5ex 10ex 5ex, width=0.483\textwidth]{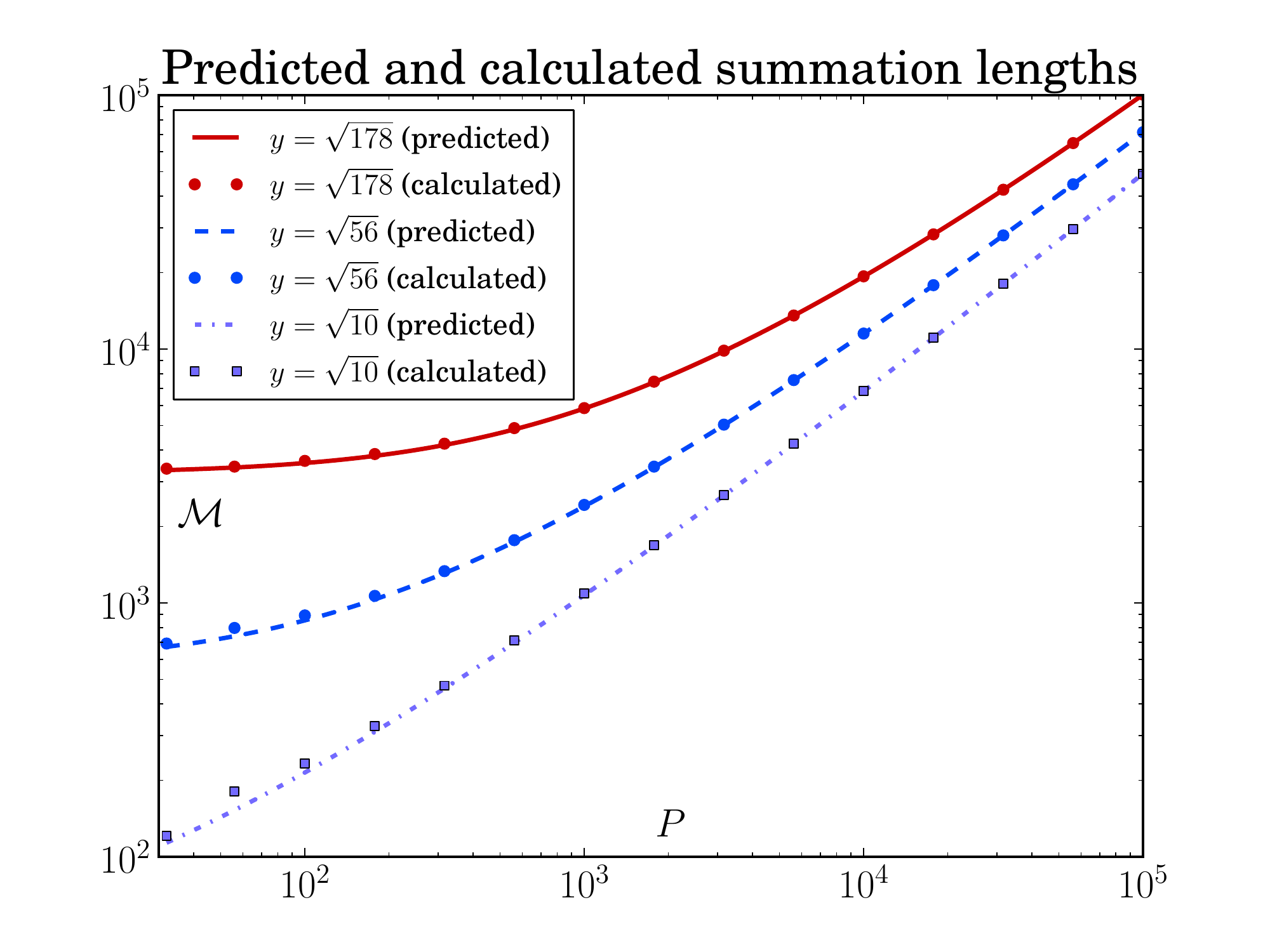}
\end{center}
\caption{This figure compares the {\em a priori\/} prediction,
based on equation~(\ref{prediction0}),
of the number of terms ${\cal M}$ which must be summed in order to evaluate
$\Psi(y)$ for $c=0$ to a desired precision $P$, with the actual number of terms
computed by the numerical routine \cite{CPC2012:AmnaKare}.
}
\label{lengthOfSums}
\end{figure}

The conclusion of this example is that for a fixed (large) $x$
we expect the largest term of the power series to be
\begin{equation}
    \mathop{\text{max}}_m \vert A_m(x) \vert \sim \text{e}^{\frac{1}{3}(x^{3/2}+3 c^2 x^{1/2})},
\end{equation}
neglecting a slowly varying prefactor.
Further, the maximum should occur at
\begin{equation}
    m \approx {\textstyle \frac{1}{2}} \left( x^{3/2} + c^2 x^{1/2} \right).
\end{equation}
Finally, estimates like equation (\ref{prediction0})
for the coefficients $a_m$ may be used to predict how many terms ${\cal M}$
we must sum to evaluate $\psi(x)$ to a given precision $P$,
based on the stopping criterium
\begin{equation}
       \vert a_{\cal M} \vert\, x^{\cal{M}} \le 10^{-P}.
\end{equation}
As can be seen in figure~\ref{lengthOfSums} the agreement with the actual
number of terms used by our evaluation routine is good,
in particular for high precision $P$. But keep in mind that a
logarithmic scale makes it easier for a comparison to look good.

\noindent
Next consider the logarithmic corrections. Including the prefactor of
equation~(\ref{CrudeWKBApproximation}) changes $S(u)$ by an amount
\begin{equation}
  \Delta S(u) = -\frac{1}{2}\log\left(\text{e}^u + c^2\right).
\end{equation}
Including the $\log(S''(u))$-term in the relation between
$S(u)$ and $S_0(u)$ changes $S_0$ by an additional amount
\begin{equation}
   \Delta S_0(u) = 
   -\frac{1}{2}\log\left(\frac{3}{4}\text{e}^{\frac{3}{2}u}+\frac{1}{4}c^2\,\text{e}^{\frac{1}{2}u}\right).
\end{equation}
For $c^2 = 0$ this changes the relation~(\ref{prediction0}) to
\begin{equation}
   \log \vert a_m \vert = \frac{1}{3}\left(2m+{5}/{2}\right)
   \left(1 -\log \left(2 m + {5}/{2}\right)\right).
   \label{prediction1}
\end{equation}
For $\vert a_m \vert$ this essentially corresponds to a factor $ m^{-5/6}$.

\begin{figure}[!t]
\begin{center}
\includegraphics[clip, trim = 10.5ex 5ex 10ex 5ex, width=0.483\textwidth]{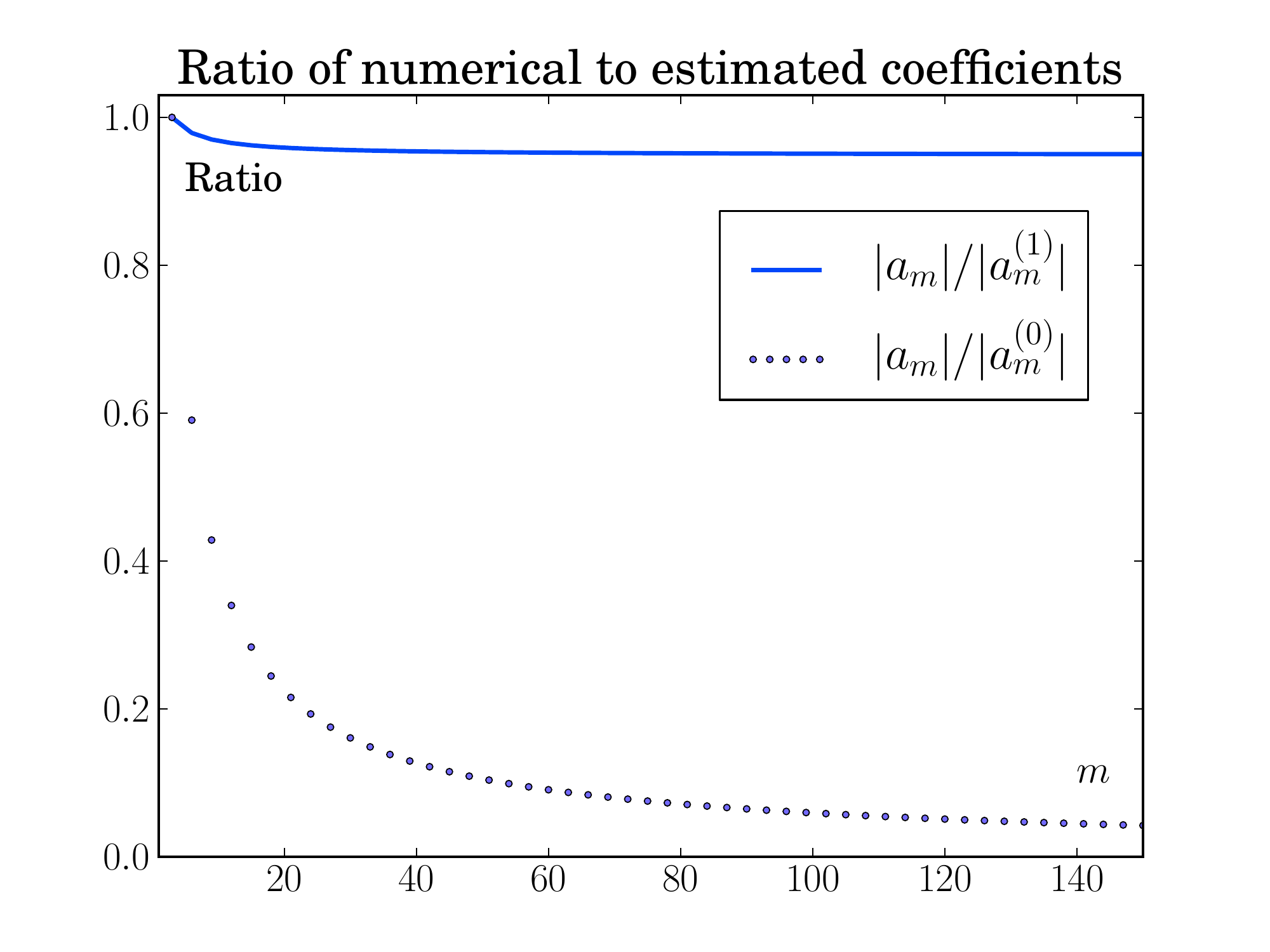}
\end{center}
\caption{This figure shows the ratio between the computed coefficients $a_m$ and the
crude prediction~(\ref{prediction0}) (labelled $\vert a^{(0)}_m \vert$) and the
logarithmically corrected prediction~(\ref{prediction1}) (labelled $\vert a^{(1)}_m \vert$).
For easy comparison we have in both cases adjusted an overall constant such that the ratio
is unity for $m=3$.   
}
\label{Ratios_a_m}
\end{figure}

\subsection{Double well oscillators}

The same procedure also work for the equation
\begin{equation}
   -\frac{\partial^2}{\partial y^2}\Psi(y) + \left(y^2- c^2\right)^2\Psi(y) = 0,
   \label{Double_well}
\end{equation}
which however is a little more challenging since the maximum value of
$\vert\Psi(y\text{e}^{\text{i}\varphi}\vert$ sometimes occur for
$\varphi \ne 0$, i.e.~for complex arguments.

For large $y$ the typical solution behaves like
\begin{equation}
     \Psi(y) \sim \text{e}^{\frac{1}{3}y^3 - c^2 y},
\end{equation}
neglecting the slowly varying prefactor. Equation (\ref{Double_well})
can be transformed to the form (\ref{ODE}) by introducing $x=y^2$, 
$\Psi(y) = \psi(x)$. Hence, with $x=y^2=\text{e}^{u}$
\begin{equation*}
     S(u) = \mathop{\text{max}}_\varphi  {\textstyle \frac{1}{3}}\text{Re} \left(\text{e}^{\frac{3}{2}(u+\text{i}\varphi)} 
       - 3 c^2 \text{e}^{\frac{1}{2}(u+\text{i}\varphi)}\right).
\end{equation*}
The maximum occurs for $\cos\frac{1}{2}\varphi = -\frac{1}{2}\left(1 + c^2\,\text{e}^{-u} \right)^{1/2}$ when
$\text{e}^{u} \ge \frac{1}{3} c^2$, and for $\cos\frac{1}{2}\varphi=-1$ otherwise.
This gives
\begin{equation}
     S(u) = \left\{\begin{array}{cc}
       {\textstyle c^2 \text{e}^{u/2} - \frac{1}{3}\text{e}^{3u/2}}&\text{for $e^u \le \frac{1}{3}c^2$,}\\[0.5ex]
       {\textstyle \frac{1}{3}} (\text{e}^{u} + c^2)^{3/2}&\text{for $e^u \ge \frac{1}{3}c^2$.}
       \end{array}
       \right.
\end{equation}
This implies that
{\footnotesize
\begin{align}
     \bar{m} &= 
     \left\{\begin{array}{lc}
         \frac{1}{2} \text{e}^{u/2}\left(c^2 - e^u\right)
         &\text{for $e^u \le \frac{1}{3}c^2$,}\\[0.5ex]
     {\textstyle \frac{1}{2}} \text{e}^{u}\,\left( \text{e}^u + c^2 \right)^{1/2}&
     \!\!\!\text{for $e^u \ge \frac{1}{3}c^2$},
     \end{array}
     \right.\nonumber\\[-2.0ex]
     \label{m_DoubleWell}\\
     \log\left(\left| a_{\bar{m}}\right|\right) &=\nonumber\\ 
     &\left\{\begin{array}{cc}
         \left(1\!-\!\frac{1}{2}u\right)c^2 \text{e}^{u/2} -\left(\frac{1}{3}-\frac{1}{2}u\right)\text{e}^{3u/2}
        &\text{for $e^u \le \frac{1}{3}c^2$,}\\[0.5ex]
     \left[\left({\textstyle \frac{1}{3}}\! -\!{\textstyle \frac{1}{2}}u \right)\text{e}^{u} 
       +{\textstyle \frac{1}{3}}c^2\right]\left(\text{e}^u + c^2\right)^{1/2}&
     \text{for $e^u \ge \frac{1}{3}c^2$}.
     \end{array}
     \right.
     \nonumber
\end{align}
}
This representation compares fairly well with the numerically generated coefficients,
as shown by the middle curve in figure~\ref{Coefficients_a_m}. However, in this case
the coefficients $a_m$ have a local oscillating behaviour. The representation
(\ref{m_DoubleWell}) should be interpreted as the local amplitude
of this oscillation.

The conclusion of this example is that we expect the largest term of the power series to be
term of the series to be
\begin{equation}
    \mathop{\text{max}}_m \vert A_m(x) \vert \sim \text{e}^{\frac{1}{3}(x + c^2)^{3/2}},
\end{equation}
neglecting the slowly varying prefactor.
Further, the maximum should occur at
\begin{equation}
    m \approx {\textstyle \frac{1}{2}} x \left( x + c^2 \right)^{1/2} 
    \approx {\textstyle \frac{1}{2}} x^{3/2} + {\textstyle \frac{1}{4}} c^2 x^{1/2}.
\end{equation}

\section{Conclusion}

As illustrated in this contribution the coefficients of Frobenius series
can be predicted to surprisingly high accuracy by use of Legendre transformations
and lowest order WKB approximations. We have also tested the validity of the
method on many other cases.

%


%

\appendices

\section*{Acknowledgment}

We thank A.~Mushtaq and I.~{\O}verb{\o} for useful discussions.
This work was supported in part by the Higher Education
Commission of Pakistan (HEC).

%

\ifCLASSOPTIONcaptionsoff
  \newpage
\fi



%

\end{document}